\documentclass[11pt,a4paper]{article}
\usepackage{amsmath,amssymb}
\usepackage{epsfig,graphicx}
\usepackage{cite}

\begin{document}

\title{
\begin{flushright}
{\normalsize Yaroslavl State University\\
             Preprint YARU-HE-06/04\\
             hep-ph/0606261} \\[20mm]
\end{flushright}
{\bf Neutrino chirality flip in a supernova\\ and the bound\\ on the neutrino magnetic moment}
}
\author{A.~V.~Kuznetsov$^a$\footnote{{\bf e-mail}: avkuzn@uniyar.ac.ru},
N.~V.~Mikheev$^{a}$\footnote{{\bf e-mail}: mikheev@uniyar.ac.ru}
\\
$^a$ \small{\em Yaroslavl State (P.G.~Demidov) University} \\
\small{\em Sovietskaya 14, 150000 Yaroslavl, Russian Federation}
}
\date{}

\maketitle

\begin{abstract}
The neutrino chirality-flip process under the conditions of the supernova core 
is investigated in detail with the plasma polarization effects in the photon 
propagator taken into account. It is shown that the contribution of the proton 
fraction of plasma is essential. New upper bounds on the neutrino magnetic moment
are obtained: 
$\mu_\nu < (0.5 - 1.1) \, \times 10^{-12} \, \mu_{\rm B}$ 
from the limit on the supernova core luminosity for $\nu_R$ emission, 
and $\mu_\nu < (0.4 - 0.6) \, \times 10^{-12} \, \mu_{\rm B}$
from the limit on the averaged time of the neutrino spin-flip. 
The best astrophysical upper bound on the neutrino magnetic moment 
is improved by the factor of 3 to 7.
\end{abstract}
 
\vfill

\begin{center}
 {\it Based on the talk presented } \\
 {\it at the XIV International Seminar Quarks'2006, } \\
 {\it St.-Petersburg, Repino, Russia, May 19-25, 2006}
\end{center}


\newpage

\section{Neutrino spin-flip in the supernova core}

Nonvanishing neutrino magnetic moment leads to various chirality-flipping 
processes when the left-handed neutrinos produced inside the supernova 
core during the collapse could change their chirality becoming sterile 
with respect to the weak interaction. These sterile neutrinos would 
escape from the core leaving no energy to explain the observed luminosity 
of the supernova. 

This process was investigated by several authors. R. Barbieri and 
R.~N. Mohapatra~\cite{Barbieri:1988} considered the neutrino spin-flip via both 
$\nu_L e^- \to \nu_R e^-$
and $\nu_L p \to \nu_R p$ scattering processes in the inner core of a supernova 
immediately after the collapse. However, they did not consider the essential 
plasma polarization 
effects in the photon propagator, and the photon dispersion was taken in a 
phenomenolical way, by inserting an {\it ad hoc} thermal mass into the vacuum 
photon propagator. 

Imposing for the $\nu_R$ luminosity $Q_{\nu_R}$ the 
upper limit of $10^{53}$ ergs/s, the authors~\cite{Barbieri:1988}
obtained the upper bound on the neutrino magnetic moment:
\begin{eqnarray}
\mu_\nu < (0.2-0.8) \times 10^{-11} \, \mu_{\rm B} \,.
\label{eq:lim_Barbi}
\end{eqnarray}

Later on, A. Ayala, J.~C. D'Olivo and M. Torres~\cite{Ayala:1999,Ayala:2000} used the 
formalism of the thermal field theory to take into account the influence of hot 
dense astrophysical plasma on the photon propagator. 
The upper bound on the neutrino magnetic moment was improved by them in the factor 
of 2:
\begin{eqnarray}
\mu_\nu < (0.1-0.4) \times 10^{-11} \, \mu_{\rm B} \,.
\label{eq:lim_Ayala}
\end{eqnarray}
However, those authors 
considered only the contribution of plasma electrons, and neglected the proton fraction.
Thus, the reason exists to reconsider the neutrino spin-flip processes 
in the supernova core more attentively. 

We will show in part, that the proton contribution into the photon propagator 
is essential, as well as the scattering on plasma protons. 

\section{Neutrino interaction with background}

\begin{figure}[htb]
\centering
\includegraphics[width=0.2\textwidth]{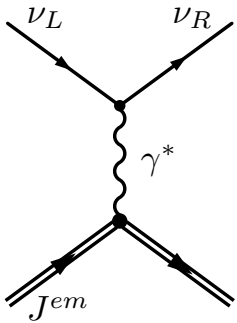}
\caption{The Feynman diagram for the neutrino spin-flip scattering via the 
intermediate plasmon $\gamma^*$ on the plasma electromagnetic current $J^{em}$.}
\label{fig:inter_plasmon}
\end{figure}

The neutrino chirality flip is caused by the scattering via the 
intermediate plasmon on the plasma electromagnetic current presented by 
electrons, $\nu_L e^- \to \nu_R e^-$,
protons, $\nu_L p \to \nu_R p$, etc. It is described by the Feynman diagram shown in 
Fig.~\ref{fig:inter_plasmon},
where $J^{em}$ is an electromagnetic current in the general 
sense, formed by different components of the medium, i.e. free 
electrons and positrons, free ions, etc. 

\def\D{\mathrm{d}} 
\def\E{\mathrm{e}}
\def\I{\mathrm{i}}

The technics of calculations of the neutrino spin-flip rate is rather standard.
The only principal point is to use the photon propagator with taking account 
of the plasma polarization effects, see e.g.~\cite{Braaten:1993}: 
\begin{eqnarray}
G_{\alpha \beta} (q) = 
\frac{- \I \,\rho_{\alpha \beta}^{(t)}}{q^2 - \Pi_t} +
\frac{- \I \,\rho_{\alpha \beta}^{(\ell)}}{q^2 - \Pi_{\ell}} \,,
\label{eq:G_alpha_beta}
\end{eqnarray}
where $\rho_{\alpha \beta}^{(t,\ell)}$ are the density 
matrices for the transversal and longitudinal plasmon, 
and $\Pi_{t,\ell}$ are the corresponding eigenvalues of the 
photon polarization tensor. 

\section{Neutrino chirality-flip rate}

Integrating the amplitude squared of the process, described by the Feynman 
diagram of Fig.~\ref{fig:inter_plasmon},
over the states of particles forming 
the electromagnetic current and over the states of the initial 
left-handed neutrinos, we obtain the rate $\Gamma (E)$ 
of creation of the right-handed 
neutrino with the fixed energy $E$.

The value $\Gamma (E)$ can be rewritten in the form 
of double integral over the energy $\omega$ and momentum 
$k \equiv |{\vec k}|$ of the virtual plasmon:  
\begin{eqnarray}
\Gamma (E) &=& \frac{\mu_\nu^2}{16\, \pi^2 \, E^2} \; 
\int\limits_0^\infty \, k^3 \, \D k 
\int\limits_{-k}^k \, \D \omega \, \theta (2 E + \omega - k) \, 
\frac{(2 E + \omega)^2}{1 - \E^{-\omega/T}}\; f_\nu (E + \omega)
\nonumber\\
&\times& 
\left[ 1 - \left( \frac{\omega}{k} \right)^2 \right]^2
\left[  \rho_\ell (\omega, k) + 
\left( 1 - \frac{k^2}{(2 E + \omega)^2} \right)\,  \rho_t (\omega, k)\right] .
\label{eq:Gamma_def}
\end{eqnarray}
Here $f_\nu (\varepsilon) 
= \left(\E^{(\varepsilon - \tilde \mu_\nu)/T} + 1\right)^{-1}$ is 
the left-handed neutrino distribution function with the chemical potential 
$\tilde \mu_\nu$, the functions $\rho_{\ell,\,t} (\omega, k)$ 
are the spectral densities of the longitudinal and transversal plasmons.

It should be noted that the factor $(1 - \E^{-\omega/T})^{-1} = 
1 + f (\omega)$ when $\omega > 0$, accounting for 
the Bose---Einstein distribution $f (\omega)$
of photons in the final state,
arises automatically in the integration over the states 
of particles forming the electromagnetic current, independently on the 
nature of particles. In the region where $\omega < 0$, this factor can be 
rewritten as $(1 - \E^{|\omega|/T})^{-1} = - f (|\omega|)$, and together with 
the change of sign of the functions $\rho_{\ell,\,t}$ when 
$\omega = -|\omega|$ is substituted, it accounts for the 
distribution of initial photons with the energy $|\omega|$, being captured 
by a neutrino. 

We note that our expression~(\ref{eq:Gamma_def}) for $\Gamma (E)$ 
is larger by the factor of 2 
than the corresponding formulas in the papers by A. Ayala 
et al.~\cite{Ayala:1999,Ayala:2000}.
On the other hand, it is in agreement, to the notations, with the rate 
obtained by P. Elmfors et al.~\cite{Elmfors:1997}.

\section{Photon dispersion}

The spectral densities $\rho_{\ell,\,t} (\omega, k)$ are defined 
by the photon polarization operator and have the form:
\begin{eqnarray}
\rho_{\ell,\,t} (\omega, k) = \frac{2 \, I_{\ell,\,t}}{(q^2 - R_{\ell,\,t})^2 
+ I_{\ell,\,t}^2}\,.
\label{eq:rho_ell_t}
\end{eqnarray}
Here, $R_{\ell,\,t}$ and $I_{\ell,\,t}$ 
are connected with the real and imaginary parts of the eigenvalues of the 
photon polarization operator 
$\Pi_{\ell,\,t} = R_{\ell,\,t} \pm \I \, I_{\ell,\,t}$, 
containing the contributions of all components of the active medium.
The functions $\Pi_{\ell,\,t}$ can be found in~\cite{Braaten:1993}. 

For the supernova conditions, the main contribution comes from the plasma 
electrons and protons:
\begin{eqnarray}
R_{\ell,\,t} \simeq R_{\ell,\,t}^{(e)} + R_{\ell,\,t}^{(p)}\,, \quad
I_{\ell,\,t} \simeq I_{\ell,\,t}^{(e)} + I_{\ell,\,t}^{(p)}\,.
\label{eq:R,I}
\end{eqnarray}
It is interesting to note that for the medium where $I_{\ell,\,t} \to 0$, 
and simultaneously the dispersion equation $q^2 - R_{\ell,\,t} = 0$
is fulfilled, the rate $\Gamma (E)$ describes the Cherenkov-like process 
with emission ($\omega > 0$) and absorption ($\omega < 0$) 
of the real plasmon (photon), investigated in part by 
W. Grimus and H. Neufeld~\cite{Grimus:1993}.

\begin{figure}[htb]
\centering
\includegraphics[width=0.85\textwidth]{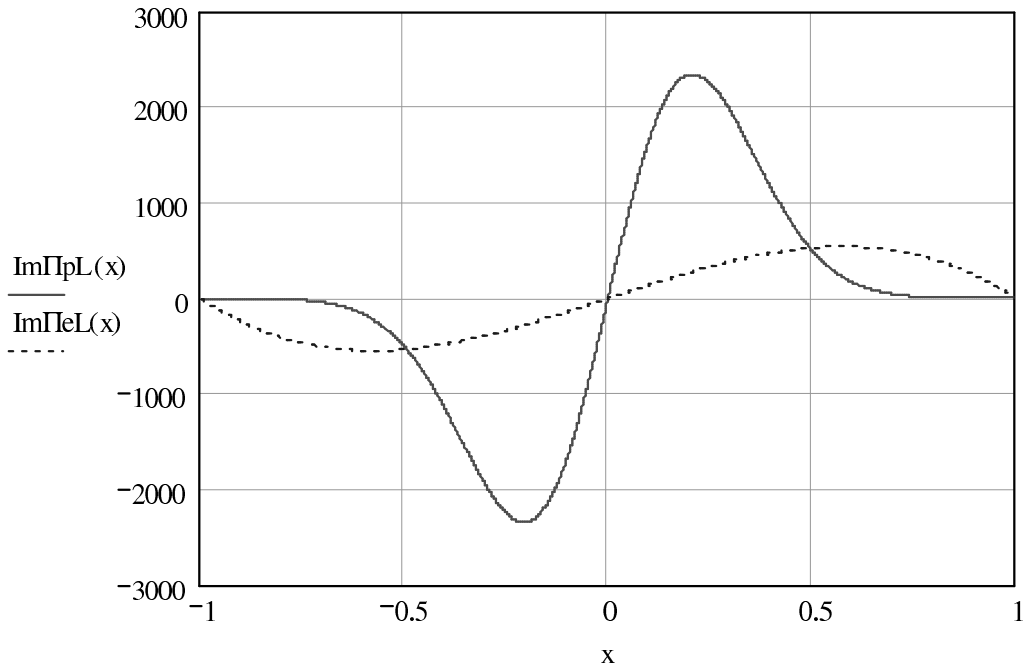}
\caption{Proton (solid line) and electron (dotted line) contributions 
to the imaginary part of $\Pi_{\ell}$.}
\label{fig:bound_fig2}
\end{figure}

\begin{figure}[htb]
\centering
\includegraphics[width=0.85\textwidth]{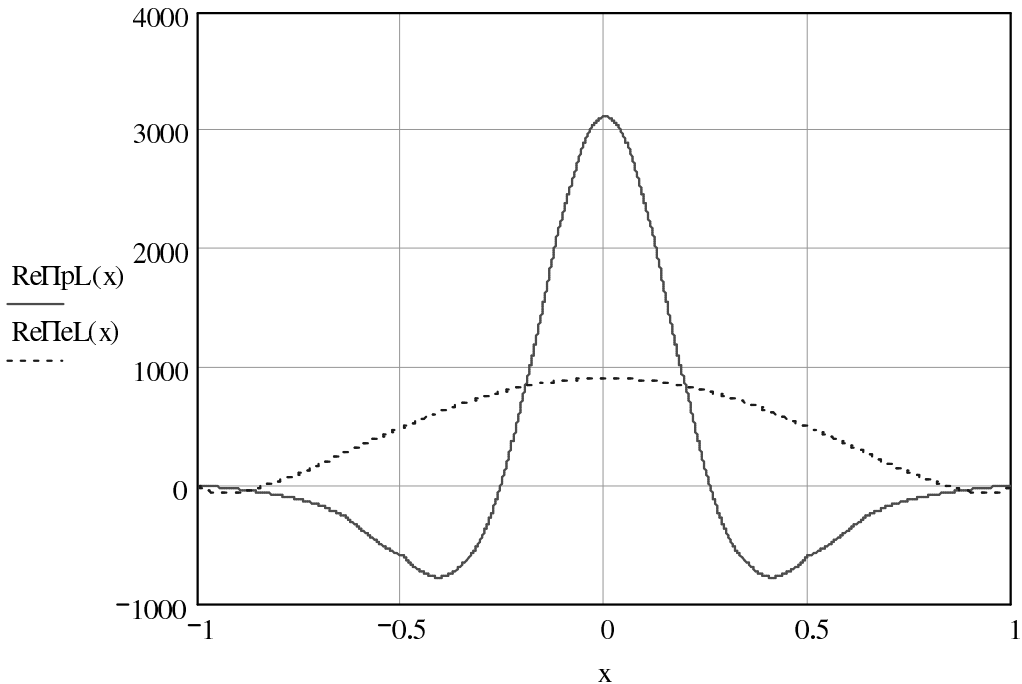}
\caption{Proton (solid line) and electron (dotted line) contributions 
to the real part of $\Pi_{\ell}$.}
\label{fig:bound_fig3}
\end{figure}

In the Figs.~\ref{fig:bound_fig2} and ~\ref{fig:bound_fig3} 
we illustrate the importance of taking into account 
the proton contribution into the eigenvalue $\Pi_{\ell}$ for the 
longitudinal plasmon. 
For the case of transversal plasmon, the proton contribution is not essential.

\section{Right-handed neutrino luminosity}
                             
\begin{figure}[htb]
\centering
\includegraphics[width=0.85\textwidth]{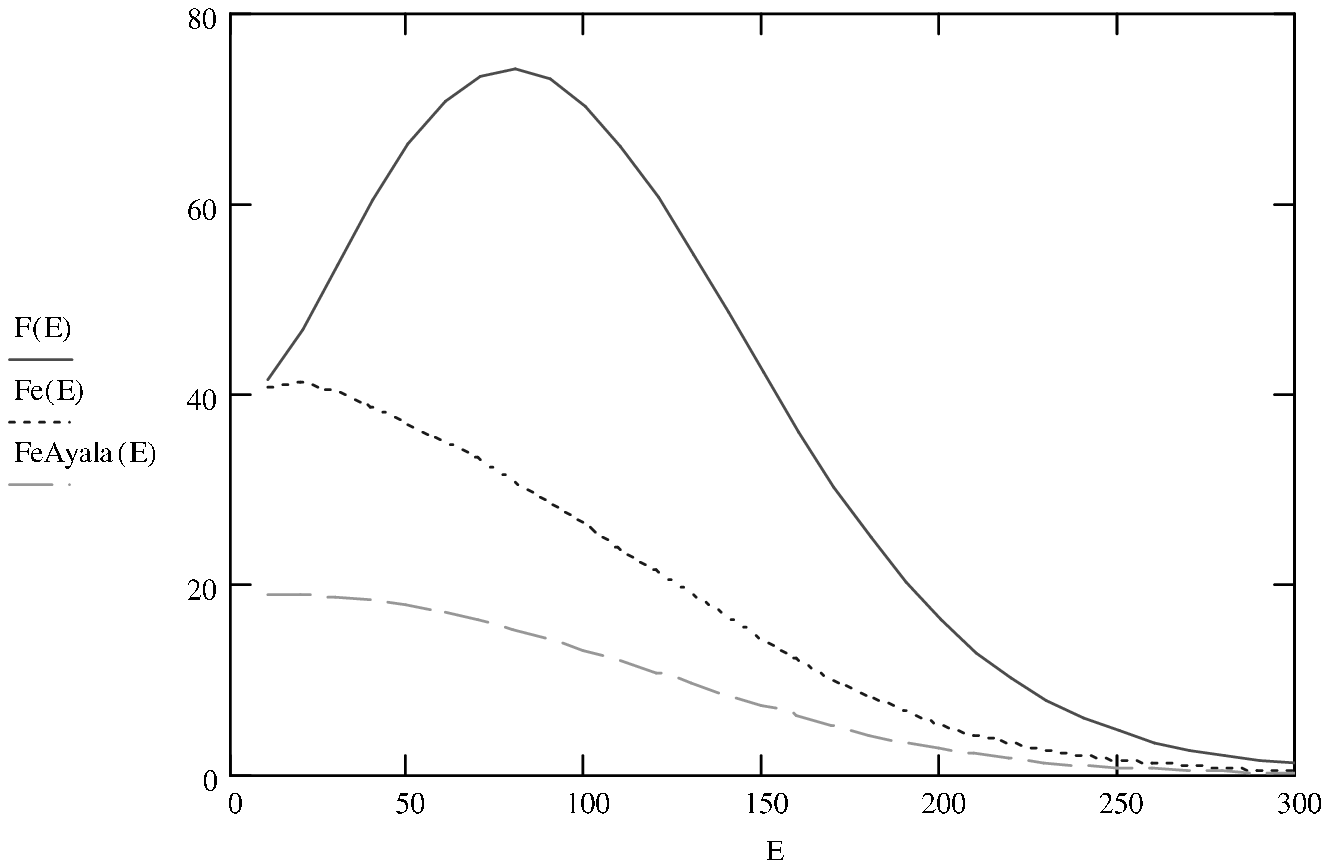}
\caption{The function $F (E)$ defining the production rate of $\nu_R$ 
with the proton contribution (solid line)
and without it (dotted line). 
The dashed line shows the result by A. Ayala et al.~\cite{Ayala:2000}.}
\label{fig:bound_fig4}
\end{figure}

The result of our numerical calculation of the neutrino chirality-flip rate
$\Gamma (E)$ is presented in the Fig.~\ref{fig:bound_fig4}. 
The plotted function $F (E)$ is defined by the expression
\begin{eqnarray}
\Gamma (E) = \frac{\mu_\nu^2\ T^3}{32\ \pi}\, F (E)\,.
\label{eq:F(E)_def}
\end{eqnarray}
It is seen from the Fig.~\ref{fig:bound_fig4} that the proton contribution is 
essential indeed. 

The supernova core luminosity for $\nu_R$ emission can be computed as
\begin{eqnarray}
Q_{\nu_R} = V\, \int \frac{\D^3 p}{(2 \pi)^3} \; E \, \Gamma (E)\,,
\label{eq:Q_def}
\end{eqnarray}
where $V$ is the plasma volume. 

For the same supernova core conditions as in the papers~\cite{Ayala:1999,Ayala:2000}
(plasma volume $V \sim 8 \times 10^{18} 
{\rm cm}^3$, temperature range $T = 30 - 60$ MeV, 
electron chemical potential range $\mu_e = 280 - 307$ MeV), 
we found
\begin{eqnarray}
Q_{\nu_R} = \left(\frac{\mu_\nu}{\mu_{\rm B}}\right)^2 (0.76 - 4.4) 
\times 10^{77} \; \mbox{ergs/s}\,.
\label{eq:Q_res}
\end{eqnarray}

Assuming that $Q_{\nu_R} < 10^{53}$ ergs/s, we obtain 
the upper limit on the neutrino magnetic moment 
\begin{eqnarray}
\mu_\nu < (0.5 - 1.1) \, \times 10^{-12} \, \mu_{\rm B}\,.
\label{eq:mu_fr_Q}
\end{eqnarray}

\section{Left-handed neutrino washing away} 

\begin{figure}[htb]
\centering
\includegraphics[width=0.85\textwidth]{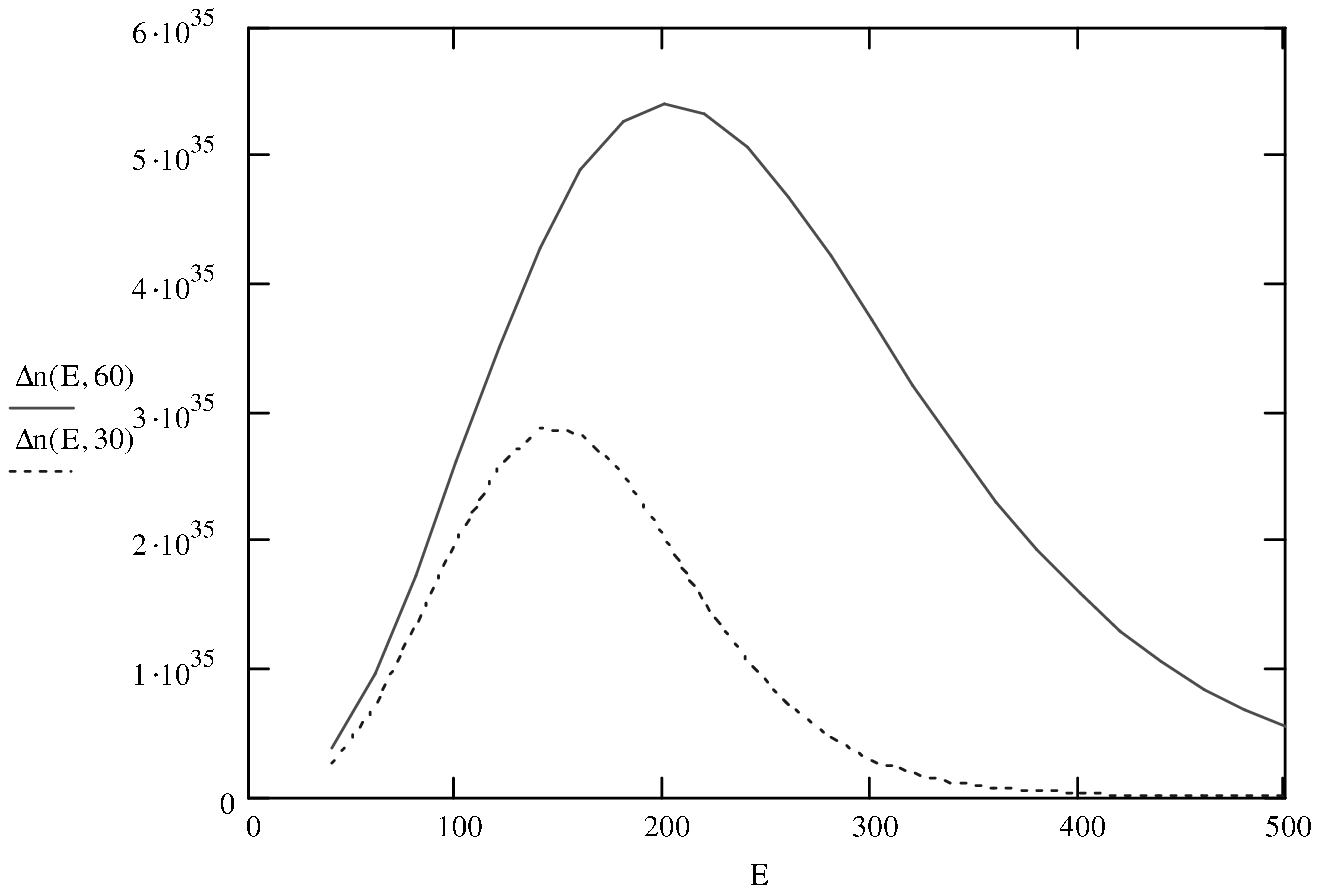}
\caption{The number of right-handed neutrinos 
(for $\mu_\nu = 10^{-12} \, \mu_{\rm B}$) 
emitted per 1 cm$^3$ 
per 1 sec per 1 MeV of the energy spectrum for 
the plasma temperature $T =$60 MeV (solid line) and for 
$T =$30 MeV (dotted line).}
\label{fig:bound_fig5}
\end{figure}

An additional method can be used to put a bound on the neutrino magnetic moment.
Together with the supernova core luminosity $Q_{\nu_R}$, a number of right-handed 
neutrinos emitted per 1 sec per 1 cm$^3$ can be defined via the rate $\Gamma (E)$ as
\begin{eqnarray}
n_{\nu_R} = \int \frac{\D^3 p}{(2 \pi)^3} \; \Gamma (E)\,.
\label{eq:n_def}
\end{eqnarray}
The right-handed neutrino energy spectrum, i.e. a number of right-handed 
neutrinos emitted per 1 sec per 1 MeV from the unit volume: 
\begin{eqnarray}
\Delta n = \frac{\D n_{\nu_R}}{\D E}
\label{eq:Delta_n_def}
\end{eqnarray}
can be also evaluated numerically. In the Fig.~\ref{fig:bound_fig5}
we show, taking for definitness $\mu_\nu = 10^{-12} \, \mu_{\rm B}$, 
the result of this calculation for two values of the plasma temperature. 

Integrating the value $\Delta n$ over all energies, one obtains 
the number of right-handed neutrinos emitted per 1 cm$^3$ 
per 1 sec. Dividing this to the initial left-handed neutrino 
number density $n_{\nu_L}$ , one can estimate the averaged time of the 
left-handed neutrino washing away, i.e. of the total 
conversion of left-handed neutrinos to right-handed neutrinos. 
For the temperature range $T = 30 - 60$ MeV, and 
for the electron chemical potential $\mu_e \sim 300$ MeV, 
we obtain 
\begin{eqnarray}
\tau \simeq \left(\frac{10^{-12}\,\mu_{\rm B}}{\mu_\nu}\right)^2 (0.14 - 0.36) 
 \; \mbox{sec}\,.
\label{eq:tau}
\end{eqnarray}
In order not to spoil the Kelvin---Helmholtz stage of the protoneutron star 
cooling ($\sim$ 10 sec), this averaged time of the neutrino spin-flip 
should exceed a few seconds. 
Taking the conservative limit $\tau >$ 1 sec, we obtain the bound 
on the neutrino magnetic moment:
\begin{eqnarray}
\mu_\nu < (0.4 - 0.6) \, \times 10^{-12} \, \mu_{\rm B}\,.
\label{eq:mu_fr_num}
\end{eqnarray}

By this means, we improve the best astrophysical upper bound  
on the neutrino magnetic moment by A. Ayala et al.~\cite{Ayala:1999}.
by the factor of 3 to 7.

\section{Conclusions}

\begin{itemize}
\item
We have investigated in detail the neutrino chirality-flip process under the 
conditions of the supernova core. The plasma polarization effects caused both by 
electrons and protons were taken into account in the photon propagator. 
The rate $\Gamma (E)$ of creation of the right-handed 
neutrino with the fixed energy $E$, the energy 
spectrum, and the luminosity have been calculated. 
\item
From the limit on the supernova core luminosity for $\nu_R$ emission, 
we have obtained the upper bound on the neutrino magnetic moment 
$
\mu_\nu < (0.5 - 1.1) \, \times 10^{-12} \, \mu_{\rm B}\,.
$ 
\item
From the limit on the averaged time of the neutrino spin-flip,
we have obtained the upper bound
$
\mu_\nu < (0.4 - 0.6) \, \times 10^{-12} \, \mu_{\rm B}\,.
$ 
\item
We have improved the best astrophysical upper bound on the neutrino magnetic moment 
by the factor of 3 to 7.
\end{itemize}

\newpage

\section*{Acknowledgements}

We express our deep gratitude to the organizers of the 
Seminar ``Quarks-2006'' for warm hospitality.

The work was supported in part 
by the Russian Foundation for Basic Research under the Grant No. 04-02-16253, 
and by the Council on Grants by the President of Russian Federation 
for the Support of Young Russian Scientists and Leading Scientific Schools of 
Russian Federation under the Grant No. NSh-6376.2006.2.


\end{document}